\pdfoutput=1

\RequirePackage{fix-cm}
\documentclass{svjour3}                     % onecolumn (standard format)
\journalname{}
  \usepackage{tipa}
  \usepackage[pdftex]{graphicx}
	\usepackage{epsfig}
	\usepackage{epstopdf}
	\usepackage{subcaption}
\newcommand\newsubcap[1]{\phantomcaption%
       \caption*{\figurename~\thefigure(\thesubfigure): #1}}

\usepackage[pdftex]{graphicx}
\usepackage{algorithmic}
\usepackage{amsmath}
\usepackage{textcomp}
\usepackage{eurosym}
\usepackage{url}
\usepackage{booktabs}

\usepackage{pgfplots}
\usepackage[utf8]{inputenc}
\pgfplotsset{every tick label/.append style={font=\small}}
\newlength\figureheight
\newlength\figurewidth

\usepgfplotslibrary{external} 
\usetikzlibrary{external}
\usepackage{float,lscape}

\usepackage{amssymb}% http://ctan.org/pkg/amssymb
\usepackage{pifont}% http://ctan.org/pkg/pifont
\newcommand{\cmark}{\ding{51}}%
\newcommand{\xmark}{\ding{55}}%

\usepackage{anysize}
\marginsize{1in}{1in}{1in}{1in}
\begin{document}

\title{Machine Learning based Soil VWC and Field Capacity
Estimation Using Low Cost Sensors %\thanks{Grants or other notes
}


\author{Idrees Zaman         \and
        Nandit Jain  \and 
        Anna F{\"o}rster %etc.
}


\institute{Idrees Zaman \at
              Sustainable Communication Networks, University of Bremen, Germany 
              %\email{iz@comnets.uni-bremen.de}           %  \\
           \and
           Nandit Jain \at
              Department of Chemical Engineering, Indian Institute of Technology, Kanpur\\
              \email{nanditjn@iitk.ac.in}
              \and
              Anna F{\"o}rster \at
               Sustainable Communication Networks, University of Bremen, Germany \\
               \email{afoerster@uni-bremen.de}\\
               \\
                \text{*} This is an extended version of I. Zaman, N. Jain and A. Förster, "Artificial Neural Network based Soil VWC and Field Capacity Estimation Using Low Cost Sensors," 2018 IFIP/IEEE International Conference on Performance Evaluation and Modeling in Wired and Wireless Networks (PEMWN), Toulouse, 2018.
}

\date{January 2021}

\maketitle

\begin{abstract}
The amount of water present in soil is measured in terms of
a parameter commonly referred to as Volumetric Water Content (VWC) and is used for determining the field capacity of
any soil. It is an important parameter accounting for ensuring
proper irrigation at plantation sites for farming as well as
for afforestation activities. The current work is an extension
to already going on research in the area of wireless underground sensor networks (WUSNs). Sensor nodes equipped
with Decagon 5TM volumetric water content (VWC) and
temperature sensor are deployed underground to understand
the properties of soil for agricultural activities. The major hindrances in the deployment of such networks over a large field
are the cost of VWC sensors and the credibility of the data
being collected by these sensors. In this paper, we analyze
the use of low-cost moisture and temperature sensors that can
either be used to estimate the VWC values and field capacity
or cross-validate the data of expensive VWC sensors before
the actual deployment. Machine learning algorithms, namely
Neural Networks and Random Forests are investigated for
predicting the VWC value from low-cost moisture sensors.
Several field experiments are carried out to examine the proposed hypothesis. The results showed that low-cost moisture
sensors can assist in estimating the VWC and field capacity
with a minor trade-off.
\end{abstract}

\section{Introduction}
\label{intro}
Substantial growth in the field of wireless sensor networks
(WSN’s) has empowered researchers to gather large amounts
of data and to reveal the sequestered information about the
nature. This information helps to understand the environment
better and guides to take appropriate decisions. This
progression in WSN leads to the evolution of its applications
that are not only limited to above ground but can be extended
to underwater wireless sensor networks (UWSN), multimedia
wireless sensor networks, mobile wireless sensor networks
and wireless underground sensor networks (WUSN). The
current work focuses on the challenges faced in agricultural
monitoring in WUSN.
WUSNs generally follow the same communication principle
as conventional WSNs but the communication range decreases
significantly because of the high permittivity of the soil.
Hence, the WUSN requires much dense deployment in
order to cover a larger area under observation. Also, the
deployment and maintenance cost of WUSN is much more
than the terrestrial WSN. There are many applications of
WUSN \cite{b1}, including, but not limited to landslide monitoring \cite{b2},
infrastructural monitoring \cite{b3}, mines disaster management \cite{b4},
agricultural monitoring \cite{b5}. The current work proposes
enhanced sustainable agricultural monitoring using WUSN.
There are several factors like soil moisture, soil temperature,
soil composition, bulk air density, volumetric water content
(VWC), pH value, conductivity etc. that determine the
suitability of a soil for agricultural and other relevant
activities. The most important factor among all mentioned
factors is the volumetric water content (VWC) of the soil.
The information about VWC at plants root zone level assists
in taking appropriate decisions for more efficient irrigation
and reforestation. The VWC is defined as the ratio of water
volume to the soil volume and is being measured using the
dielectric permittivity of the soil. Soil VWC information is of
great use for providing irrigation solutions to the farmers as
well as its integration with wireless sensor networks allows
monitoring of the remote agricultural places and reforestation
projects. A wide range of expensive VWC sensors are
available for monitoring the water content of a soil. The
cost of VWC sensors become a bottleneck for the farmers
to opt the techniques of smart agriculture to increase their
crops productivity. The current work strives to address the
aforementioned issues regarding soil monitoring.\\
The current work is a part of the cooperation between the Sustainable Communication Networks group at the University of Bremen in Germany and the ReviTec \cite{b6},\cite{b7} project from the Center of Environmental Research and Sustainable Technology (UFT) at the University of Bremen. The agricultural fields in Ngaoundere, Cameroon where the project is aimed to be deployed are often located in remote areas far away from settlements. In such regions, vandalism is quite frequent from anti social elements. Also, there is a risk of sensors and experiment being harmed by the farmer as well as  wild animals. Given such circumstances, it is advisable to use underground sensors with no visible parts above the ground. \\
The aim of this work is to help ecologists to evaluate the performance of their revitalization techniques and common farmers to adopt soil monitoring techniques to better understand the nature of their agricultural fields. The WUSN also helps farmers to pursue their normal agricultural field operations without any hindrance as all the sensor nodes are buried underground. The current work will also help in a much denser deployment of the sensors which otherwise is not possible because of the expensive volumetric water content sensors.

\section{Related Work}

Mainly, there are two methods for measuring the VWC of the soil i.e.\ aerial sensing and underground sensing. The subsequent sections will highlight the recent work carried out in both the approaches along with their advantages and shortcomings.

\subsection{Aerial Sensing}
The process of aerial sensing requires satellite imaging or
high altitude aircraft imaging to get viable information for the
ground soil. For the purpose of VWC estimation, it is generally done through imaging or electromagnetic (EM) waves.
In this direction, NASA launched Soil Moisture Active Passive (SMAP)\cite{b8} in 2015, that is an American environmental
research satellite. SMAP has been designed to measure soil’s
moisture content over a three-year period, every 2-3 days. The
SMAP observatory consists of a spacecraft mounted with instruments in a near-polar, Sun-synchronous orbit. The SMAP
consists of a radiometer (passive) instrument and a synthetic
aperture radar (active) instrument operating with multiple
polarizations in the L-band range. The project is aimed at
developing global maps of soil moisture for areas not covered
by water or snow.
In \cite{b9}, authors used Bayesian Artificial
Neural Network based approach has been proposed to extract soil’s VWC information from high resolution imaging
drone through an autonomous unmanned aerial system (UAS),
named AggieAir, equipped with visual, near-infrared, and
thermal cameras. The imaging was done at 0.15 m resolution
for the visible and near infrared bands and 0.6 m resolution
for the thermal infrared band. Multiple indices like red, green,
blue, NIR, thermal, NDVI, VCI, EVI, VHI, and field capacity
were extracted from the images to train the models. The study
area was the 0.34 km2 agricultural field located in Scipio,
Millard County, in central Utah. Intensive ground sampling
was carried out to detect the ground truth. The model could
estimate root zone’s VWC with a good accuracy (RMSE =
0.05).\\
In \cite{b10}, the authors used the satellite data to obtain
VWC data. The authors use Global Vegetation Moisture Index
(GVMI) to retrieve vegetation water content. The index was
created based on an analytical method using radiative transfer
models at leaf, canopy, and atmospheric levels, with an emphasis to maximise sensitivity to vegetation water content and
to minimize sensitivity to other factors such as atmospheric
perturbations and angular effects. The index was validated
on areas covered with four different vegetation types, representing a vegetation gradient from open herbaceous (Shrub
Steppe) to the densely vegetated formation (Savannah Woodland). A linear correlation coefficient of 0.93 was observed
between GVWI derived VWC and field measurements.
The aerial sensing method is generally quite expensive, given
that satellites or aircrafts and high definition cameras are involved. Usually, the aforementioned methods are used for
estimating the VWC profile over very large areas and are not
feasible enough to be used by farmers for their agricultural
fields.

\subsection{Underground Sensing}

In the recent past, researchers have proposed several techniques to measure the VWC of large agricultural fields using underground sensing. In G. Shan et al. \cite{b11}, an underground mobile dielectric sensor is used to measure the VWC of the tomato crop root zone in an experimental field sizing 240m$^{2}$. The VWC is logged twice a day continuously for a period of three months. A stepper and dc motor are used to move the dielectric sensor in steps of 5cm in a buried PVC tube of length 4m. The metal electrodes of the embedded sensor are firmly adjoined to the PVC tube. The probes of the dielectric sensor generate a fringing field that passes through the plastic PVC tube into the soil and measures the dielectric permittivity of the soil.  The mobile dielectric sensor allows to cover a larger area with a single sensor.  The data collected gives useful insight in soil and plant water losses linked with evapotranspiration. The data collected through the sensor is also cross-validated against the estimated model of the root zone soil. Such setup requires high deployment and maintenance cost.

In Mittelbach et al. \cite{b12}, the authors evaluate the performance of  commercially available expensive VWC sensors from Decagon Devices. The work involved field measurements performed at two sites with different soil texture and calibration techniques in Switzerland covering more than a year of parallel measurements in several depths down to 120 cm. The authors observed that the measurement accuracy of the 10HS sensor decreased considerably with increasing VWC and became insensitive to variations of VWC above 40\%, whereas the 5TM sensor showed better sensitivity to higher VWC values. To the best of our knowledge, no one has used the low cost soil monitoring sensors for estimating the VWC value. \\

\section{Soil Sensing}
The volumetric water content (VWC) is one of the important parameter to understand the feasibility of any soil for agricultural activities. The VWC is then used to derive the field capacity of any soil, which is another parameter to analyze the suitabitly of any soil for agriculture. However, measuring of VWC require complex methods i.e. time domain reflectometry (TDR) and frequency domain reflectometry (FDR). The sensing modules based on TDR and FDR are usually quite expensive i.e. 5TM Decagon sensor \cite{b13}. Therefore, potential low cost sensors that can estimate the VWC in place of TDR and FDR based methods are analyzed.

\subsection{Field Capacity}

Soil is normally composed of three types of particles, i.e.\ sand, silt and clay. The proportion of these particles in the soil dictate the suitability of the soil for agricultural activities. The porosity of soil depends on the pore size and pore density, that determines the field capacity of any soil. Field capacity \cite{b14} is the remaining VWC in the soil after two to three days of a rainfall, when most of the water is drained by the gravity. While determining the field capacity, it is assumed that no water is used by plant roots or evaporated. Therefore it is preferable to estimate the field capacity in winter so that the chances of water evaporation are minimum. Field capacity depends highly on the composition of particles in the soil. For example the sandy soil has a much smaller field capacity than the soil with clay particles, as the water gets drained much quickly in sandy soil than the soil with clay particles.

\subsection{Need for Alternate Sensors}

The soil's VWC is a localized observation, and multiple nodes are needed to analyze large areas. The cost of a single 5TM VWC sensor ranges from Euro 180 to Euro 220, depending on the vendor. It is not a practical solution to deploy a large number of such expensive sensors in a large agricultural field. Therefore, an investigation into alternate solutions is required to reduce the deployment cost for agricultural monitoring. 
Table \ref{tab:table1} shows all the soil sensing sensors and the subsequent section explains the principal of these sensors in detail. 

\begin{table}[h!]
  \centering
  \caption{Sensor Details.}
  \label{tab:table1}
  \begin{tabular}{ccc}
    \toprule
    Sensor Name & Measurement & Price\\
    \midrule
    DS18S20 (OneWire)\cite{b15} & Temperature & \euro 15.5\cite{b16}\\
    
    SHT10\cite{b17} & Temp \& Humidity & \euro 54\cite{b18}\\
    
    YL-69 & Moisture & \euro 1.3\cite{b19}\\
    
    SEN13322\cite{b20} & Moisture & \euro 4.9\cite{b21}\\
    
    5TM Decagon\cite{b13} & VWC & \euro 180\cite{b23} \\
    
    \bottomrule
  \end{tabular}
\end{table}

\label{sensors}
\subsection{5TM Decagon Sensor}
5TM decagon is one of the most reliable and expensive sensors used for VWC monitoring. The sensor uses the time domain reflectometry (TDR) method for measuring the dielectric permittivity of the soil. It measures the time taken by an electromagnetic pulse to go through a transmission line and return back. This time taken is related to the dielectric permittivity $\kappa$ using Equation \ref{timeeq} \cite{b24}.

\begin{equation}\label{timeeq}
\kappa = (\frac{tc}{2L})^2
\end{equation}

where, $L$ is the length of the transmission line, $c$ is the speed of light and $t$ is the time taken by the electromagnetic pulse. 

 The Topp's equation \cite{b25}  is: \[VWC = 4.3\cdot10^{-6}*\varepsilon_\mathrm{a}^3-5.5\cdot10^{-4}*\varepsilon_\mathrm{a}^2+2.92\cdot10^{-2}*\varepsilon_\mathrm{a}-5.3\cdot10^{-2}\] where $\varepsilon_\mathrm{a}$ is the dielectric permittivity value reported by the 5TM sensor. The VWC values obtained by substituting the dielectric permittivity values are considered as ground truth. 

\subsection{Alternate Sensors}
 YL-69 is a low cost sensor used to measure the moisture of the soil.  The sensor has both digital and analog outputs. The sensor measures the soil resistivity by passing current between its two prongs. The increase in moisture will offer less resistance, resulting in higher current and eventually higher values at the output of the sensor.  

 SEN-13322 is also a moisture sensor and follows the same measurement principle as the YL-69 moisture sensor. The SEN-13322 only gives an analog output which can be used to calculate the moisture in the soil.

The SHT10 sensor measures the humidity and temperature. The sensor has an inbuilt 14 bit A/D converter for temperature measurements and 12 bit A/D converter for humidity measurements. The sensor has a serial interface that can be connected to any microcontroller. 

DS18S20 sensor is used to measure the temperature of the soil. The use of a standalone temperature sensor allows to cross validate the data from other sensors used for soil monitoring. DS18S20 also has a built-in 9 bit A/D converter and uses a 1-wire protocol to communicate with any microcontroller.

\section{Experimentation Setup}

In order to cross validate the proposed hypothesis, an experiment was performed at the ReviTec  site \cite{b6}. Although the experimental field is only composed of sand particles but any soil type can be used for the experiment as it will only affect the slope of rising and falling values in VWC. The experiment was carried out from June 29, 2017, to July 10, 2017 when the weather conditions were feasible for the field capacity estimation. A total of 7262 observations were recorded. \\
 The low cost sensors as well as a Decagon 5TM-VWC-Temperature sensor mentioned in Section \ref{sensors} are attached to our MoleNet\cite{b27} sensor node and are placed inside a waterproof box. 
 
\begin{figure}[ht]
\begin{center}
  \includegraphics[width=0.96\textwidth]{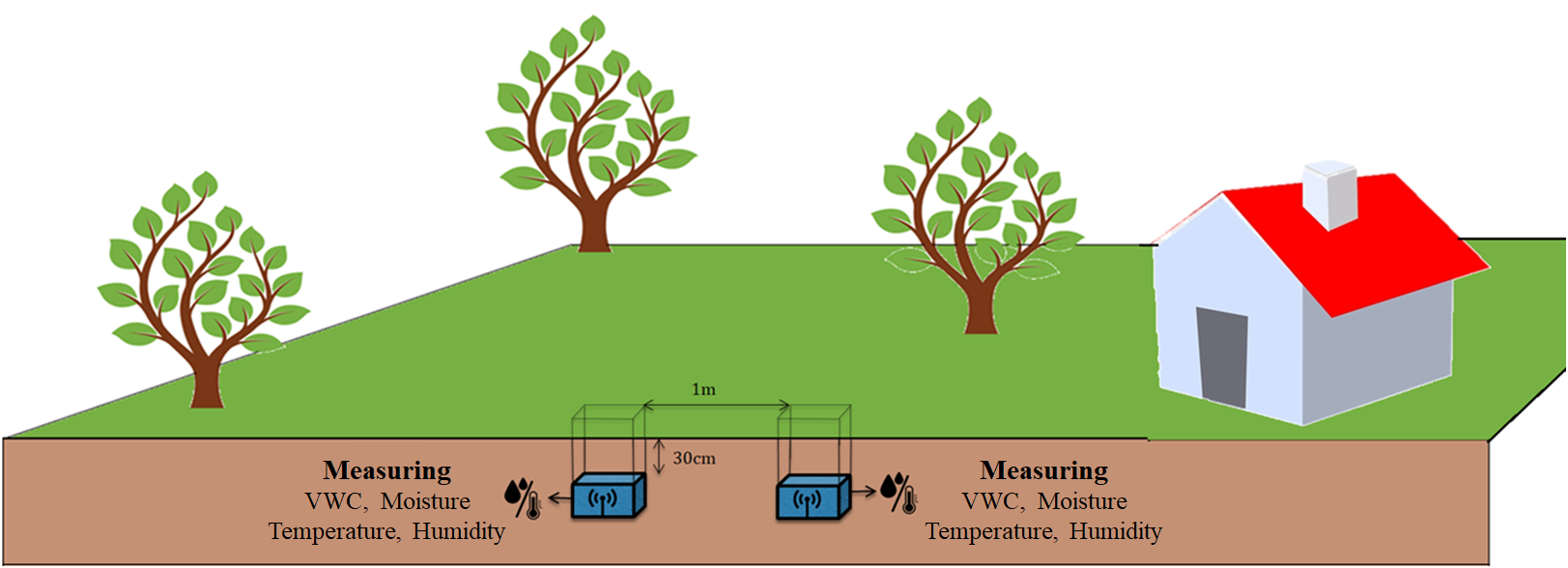}
  \caption{Experimental setup for gathering soil data }
\end{center}
  \label{fig:molenet}
\end{figure}% 
 
The sensor node samples the data from all the sensors every two minutes and sends the information to the base station.

 Two MoleNet sensor nodes equipped with all the mentioned sensors are buried at a depth of 30cm. The internodal distance between the buried nodes is kept relatively small (1m) to have similar kind of data to be used for the ground truth later on. An aboveground base station is used for receiving the packets from buried sensor nodes.

\subsection{Observations}
 
\begin{figure}[h]
\centering
\begin{subfigure}{.5\textwidth}
  \centering
  \includegraphics[width=.95\linewidth]{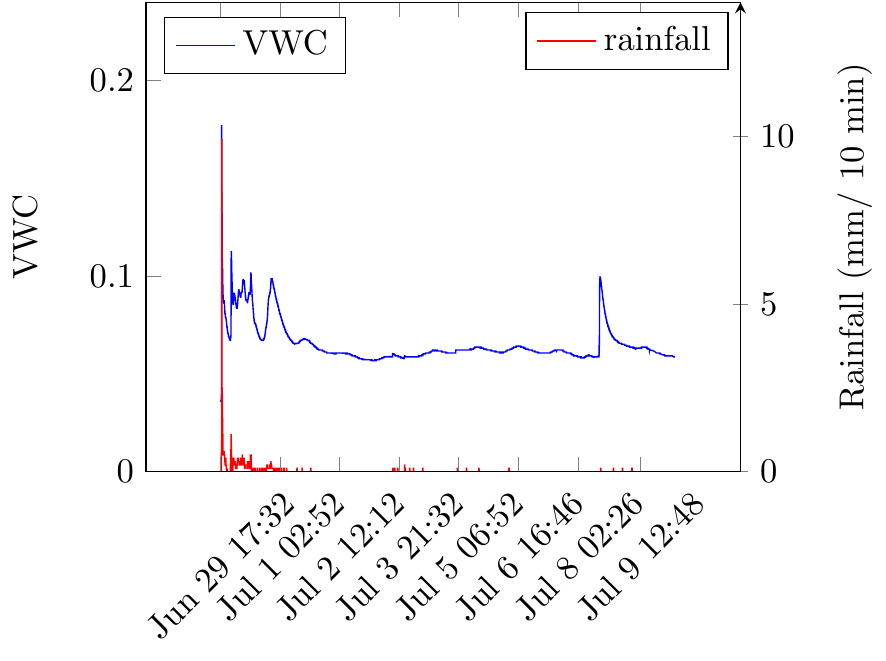}
  \newsubcap{5TM VWC Sensor and rainfall data}
  \label{VWC_rain}
\end{subfigure}%
\begin{subfigure}{.5\textwidth}
  \centering
  \includegraphics[width=.95\linewidth]{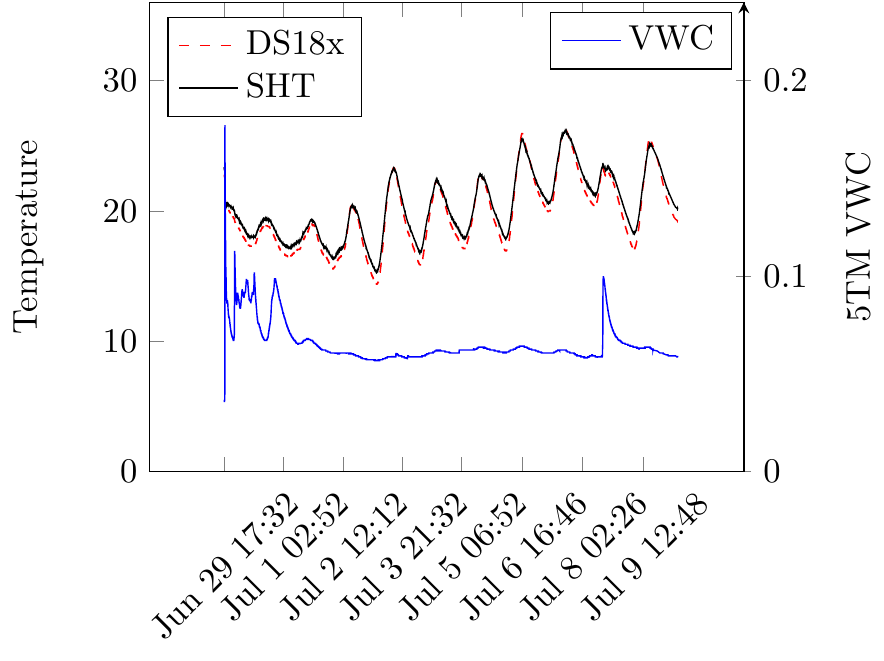}
  \newsubcap{Temperature and 5TM VWC Sensor data}
  \label{temp_rain}
\end{subfigure}
%\caption{A figure with two subfigures}
\label{fig:test1}
\end{figure}

During the experimentation period, it rained multiple times. The actual rainfall data for the experimentation period is taken from a weather station, located 1.5km away from the experimental field. Figure \ref{VWC_rain} shows the data from the VWC sensor against the rainfall data during the experiment. The rainfalls and the dry period in between allowed precise field capacity estimation. The initial peaks in the graph validate the heavy rainfall after the initial deployment of the sensor nodes. The last peak in VWC data does not perfectly reflect the amount of rain shown in the graph. One possible reason can be the location of the weather station that is approximately 1.5km away from the experimental field. \\

\begin{figure}[h]
\centering
\begin{subfigure}{.5\textwidth}
  \centering
  \includegraphics[width=.95\linewidth]{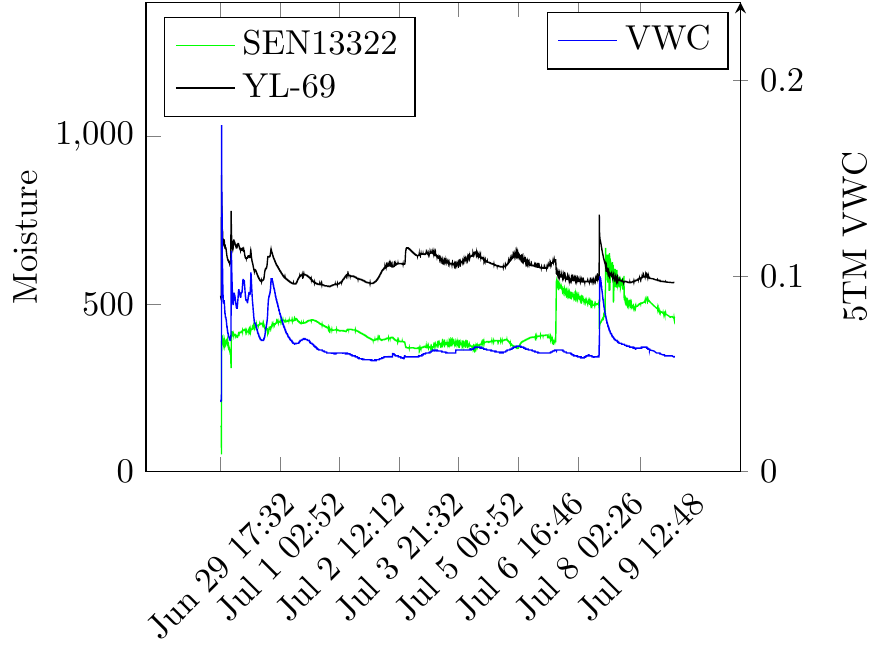}
  \newsubcap{Moisture and 5TM VWC Sensor data}
\label{moist5tm}
\end{subfigure}%
\begin{subfigure}{.5\textwidth}
  \centering
  \includegraphics[width=.95\linewidth]{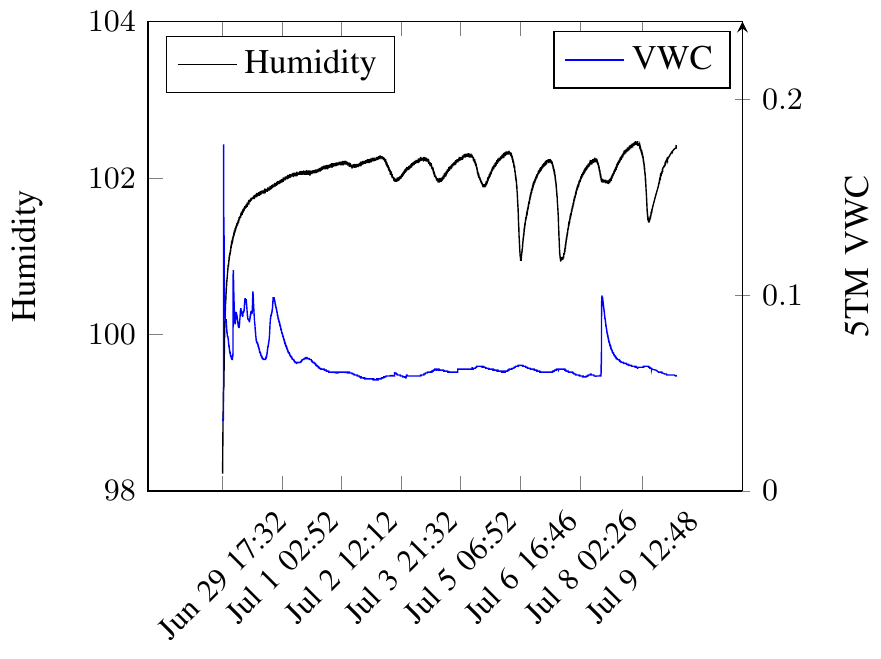}
  \newsubcap{SHT10 Humidity and 5TM VWC Sensor data}
\label{humidity}
\end{subfigure}
%\caption{A figure with two subfigures}
\label{fig:test}
\end{figure}

 Figures \ref{temp_rain}, \ref{moist5tm} and \ref{humidity}  show the temperature, moisture and humidity data against the VWC data respectively. The reason for the dissimilar values of the deployed moisture sensors in Figure \ref{moist5tm} is the different measuring principle of both sensors. Therefore raw sensor values are used instead of converting them into respective moisture values.  The data from all the low cost sensors is then used for regression modeling.

\setlength\figureheight{2.5in}
\setlength\figurewidth{3in}

\subsection{Estimated Field Capacity}
Figure \ref{VWC_rain} shows that, after the first rainfall, there was no rain till July 2nd, 2017. Although the experiments are performed in summer, Figure \ref{temp_rain} shows that the temperature after the rain remained below 20°C until the next rainfall and also there is not much vegetation on the experimental field. Therefore, it can be assumed that most of the water is drained by gravity. The field capacity of the experimental field is approximately 5.5\%. \\
In the subsequent section, the data collected from all low cost sensors is used to predict the VWC data and eventually the field capacity of the experimental field.

\section{Model Fitting}

Regression based model fitting is used to find a relationship between a response variable i.e. volumetric water content (VWC) against one or more predictor variables i.e. temperature, humidity and moisture sensor values. In current work, Support Vector Machine Regression (SVR), Random Forests and Artificial Neural Networks (ANNs) are used for regression based model fitting. K-fold cross validation is used for the validation of the predicted model over the test data. Root mean square error (RMSE), mean absolute error and Pearsons Coefficient R parameters are used. The subsequent section will briefly explain these performance evaluation parameters.

\subsection{Performance Evaluation Parameters}
Root mean square error (RMSE) and mean absolute error (MAE) are the two most used parameters to determine the accuracy of the predicted model to the actual model.  RMSE calculates the square root of the average of square differences between the predicted data and actual data. Equation \ref{eqn-rmse}  shows how to calculate the RMSE.

\begin{equation}
RMSE = \sqrt{\frac{1}{n} \sum_{i=1}^{n}\left (x_{i}  -  \hat{x_{i}} \right )^{2}}
\label{eqn-rmse}
\end{equation}
where $x_{i}$ is the actual data and $\hat{x_{i}}$ is the estimated data. Whereas, MAE calculates the average magnitude of the error in the estimated dataset. Equation \ref{eqn-mae}  shows how to calculate the MAE.

\begin{equation}
MAE = \frac{1}{n} \sum_{i=1}^{n} | x_{i}  -  \hat{x_{i}}  |  
\label{eqn-mae}
\end{equation}

Pearson correlation coefficient (R) measures the correlation between two datasets, in our case the actual dataset and the predicted dataset. A value of R equal to 1 indicates a strong correlation between the actual and predicted dataset, while a value equal to zero indicates no correlation between two datasets. Equation \ref{eqn-r} shows how to calculate R.
\begin{equation}
R = \frac{n \left ( \sum x \hat{x} \right ) - \left ( \sum x  \right )\left ( \sum \hat{x}  \right )}{\sqrt{\left [ n  \sum x^{2} - \left ( \sum x \right )^{2}  \right ]\left [ n  \sum \hat{x}^{2} - \left ( \sum \hat{x} \right )^{2}  \right ]}} 
\label{eqn-r}
\end{equation}

\subsection{Support Vector Regression (SVR)}
Support vector regressors (SVRs) are a class of machine learning algorithms which is based on the functions of the kernel based learning. The Support Vector Regression (SVR) follows the same principles for the regression as the support vector machine (SVM) does for the classification of the dataset. However, in case of regression the possibilities are infinite in the output class. Therefore, a tolerance error (epsilon) is introduced in order to approximate the SVM. SVR strives to minimize the generalization error bound rather than minimizing the observed training error which results in better generalized performance \cite{b28}.\\
 SVR uses different kernel functions i.e.\ polynomial, radial base function (RBF)\cite{b29} and sigmoid. Equation \ref{eqn:svr} shows the gausian radial basis function. The kernel functions in SVR transform the data  into a higher dimensional feature space in order to perform the linear separation.\\ 
 
 \begin{equation}
 K(x_{i},\hat{x_{i}})= exp \big(  -\frac{{|| x_{i} - \hat{x_{i}} ||}^{2}}{2\sigma^{2}} \big )
 \label{eqn:svr}
   \end{equation}
%\subsubsection{VWC and Field Capacity Estimation Using SVRs}

\subsubsection*{\textbf{Results}}
Support vector machine regressor (SVR) uses the input from all the low cost senors i.e.\ DS18S20 temperature, SHT10 humidity, SHT10 temperature, YL-69 moisture and SEN13322 moisture sensor, in order to predict the volumetric water content (VWC) and field capacity of the soil under observation. Radial basis function kernel is used to train the SVM.

\begin{figure}[h]
	\begin{center}
		\includegraphics[scale=0.9]{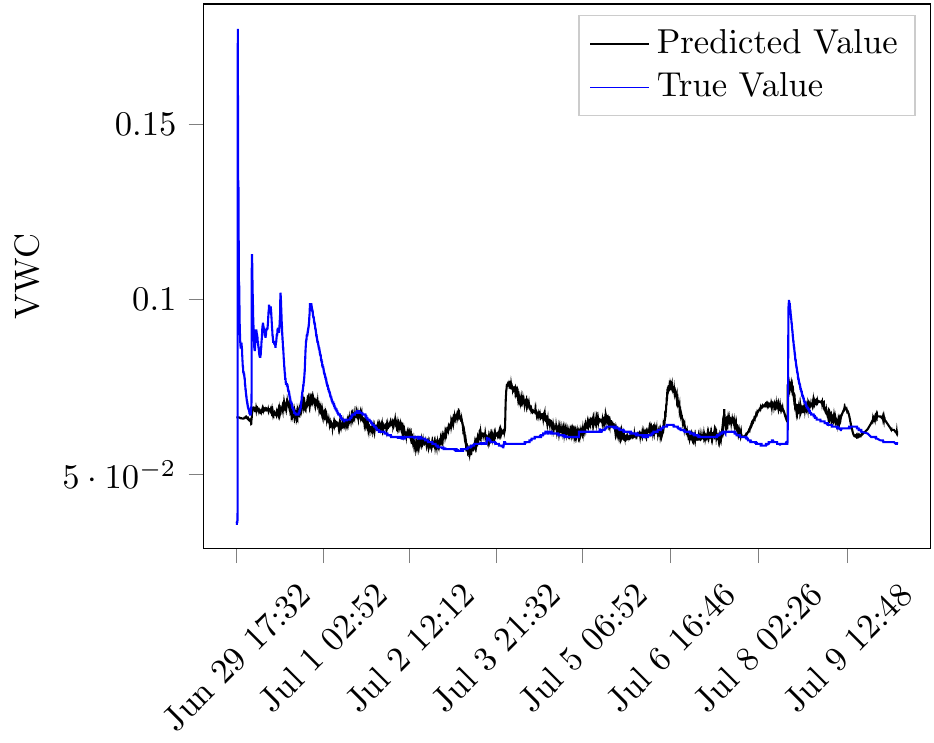}
	\end{center}
\caption{SVR based VWC Prediction using all inputs}
\label{fig:svr}
\end{figure}

The prediction results from the Support Vector Machine based regression modeling are shown in Table \ref{table:svr}. The predicted values in Figure \ref{fig:svr} show significant variation even during the stable phase of the VWC data. Therefore, estimation of field capacity is not possible using Random forests. \\

\begin{table}[h]
\centering
\caption{Performance indicators for SVR based VWC prediction}
\label{table:svr}
\begin{tabular}{|l|l|}
\hline
\textbf{Sensor Cost}    & 75.7€                                                                                                                         \\ \hline
\textbf{Algorithm}      & SVR \\ \hline
\textbf{Training Input} & \begin{tabular}[c]{@{}l@{}}DS18S20 and SHT10 (Temperature), \\ YL-69 and SEN13322 (Moisture),\\ SHT10 (Humidity)\end{tabular} \\ \hline
\textbf{RMSE}           & 0.00927354
 \\ \hline
\textbf{MAE}            & 0.00564785
 \\ \hline
\textbf{Pearson's R}    & 0.40901482
 \\ \hline
\end{tabular}
\end{table}

\subsection{Random Forests}
Random Forest is an ensemble based machine learning algorithm and is a combination of multiple decorrelated decision trees. 

 In case of the regression based tree, the target is a real number instead of a class and the leaf node corresponds to a real number. A Random Forest model trains multiple decision trees simultaneously where each training strives to minimize a cost function i.e.\ mean square error or mean absolute error in case of regression modeling.

\begin{figure}[h]
	\begin{center}
		\includegraphics[scale=1.1]{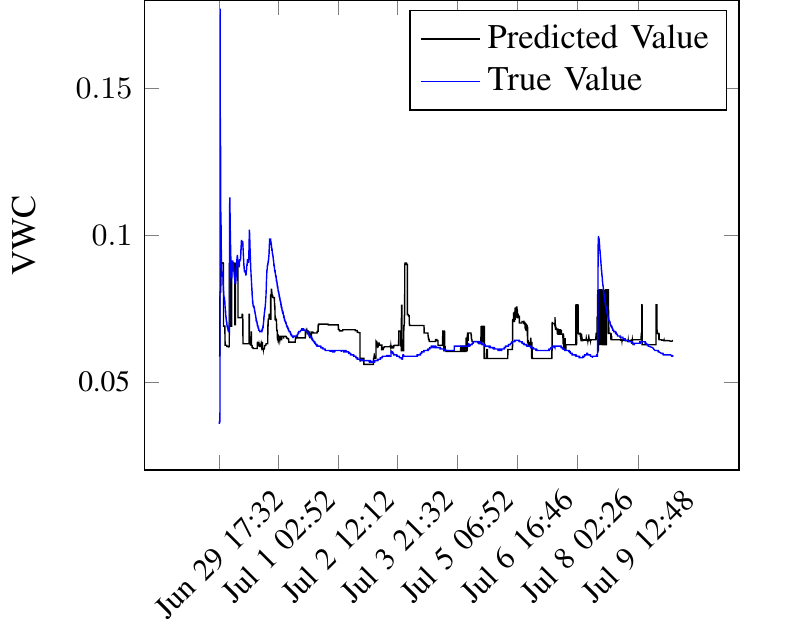}
	\end{center}
\caption{Random forest based VWC Prediction using all inputs}
\label{rf_all}
\end{figure}

For the purpose of training Random Forest model, the Scikit-Learn \cite{b30} learning method is used. Initially all the input variables i.e.\ DS18S20 temperature, SHT10 humidity, SHT10 temperature, YL-69 moisture and SEN13322 moisture are used for training the model. All the inputs are normalized and the mean and standard deviation of all the parameters of the training and testing sets are calculated. Table \ref{configRandom} shows the configuration parameters used for training the Random Forest based regression model.

\begin{table}[h]
\centering
\caption{Configuration Parameters for Random Forest}
\label{configRandom}
\begin{tabular}{|c|c|}
\hline
\textbf{Parameter}   & \textbf{Value} \\ \hline
Number of Estimators & 24             \\ \hline
Number of Leaf Nodes & 30             \\ \hline
Maximum Depth        & 7              \\ \hline
\end{tabular}
\end{table}
Root mean squared error (RMSE), mean absolute error (MAE) and Pearson's correlation coefficient (R) are used as the performance evaluation parameters for the learned model. In order to avoid overfitting of the learned model over the training dataset, five fold cross validation method is used. The entire dataset is divided into 5 equal subsets. One subset is held out while the remaining four subsets are used  to train the regression model and the resultant model is used to make predictions over the held out subset. The whole process is repeated five times, once for each of the five subsets.

\subsubsection*{\textbf{Results}}
 The prediction results from the Random Forest based regression modeling are shown in Table \ref{table:rf1}. The predicted values in Figure \ref{rf_all} show significant variation (even more than SVR)  even during the stable phase of the VWC data. Therefore, estimation of field capacity is very difficult using Random forests. \\
In case of decision trees and Random Forests, the predictions are a limited number of discrete values, depending on the number of leaf nodes. Therefore, the only axis aligned separation is possible with Random Forests, as visible from the horizontal lines in the 'Predicted Value' curve. For improved regression modeling using Random Forests, the number of leaf nodes should be increased manifolds which further would increase the complexity of the model significantly.

\begin{table}[h]
\centering
\caption{Performance indicators for Random Forest based VWC prediction}
\label{table:rf1}
\begin{tabular}{|l|l|}
\hline
\textbf{Sensor Cost}    & 75.7€                                                                                                                         \\ \hline
\textbf{Algorithm}      & Random Forest                                                                                                                 \\ \hline
\textbf{Training Input} & \begin{tabular}[c]{@{}l@{}}DS18S20 and SHT10 (Temperature), \\ YL-69 and SEN13322 (Moisture),\\ SHT10 (Humidity)\end{tabular} \\ \hline
\textbf{RMSE}           & 0.00907407                                                                                                                    \\ \hline
\textbf{MAE}            & 0.00621031                                                                                                                    \\ \hline
\textbf{Pearson's R}    & 0.47870054                                                                                                                    \\ \hline
\end{tabular}
\end{table}

\subsection{Gradient Boosting Regression}
In order to improve the prediction results of the random forests over the stable area of the volumetric water content, gradient boosting regression technique is used. Gradient boosting is an ensemble technique in which predictions are made based on the weak predictions of a regression technique i.e. decision trees. Table \ref{configGradient} shows the configuration parameters for our gradient boosting regression model.

\begin{table}[h]
\centering
\caption{Configuration Parameters for Gradient Boosting Regression}
\label{configGradient}
\begin{tabular}{|c|c|}
\hline
\textbf{Parameter}   & \textbf{Value} \\ \hline
Random State & 0             \\ \hline
Number of Estimators & 100             \\ \hline
Maximum Leaf Nodes        & 25              \\ \hline
Maximum Depth        & 3              \\ \hline
\end{tabular}
\end{table}

\subsubsection*{\textbf{Results}}

\begin{figure}[h]
	\begin{center}
		\includegraphics[scale=0.9]{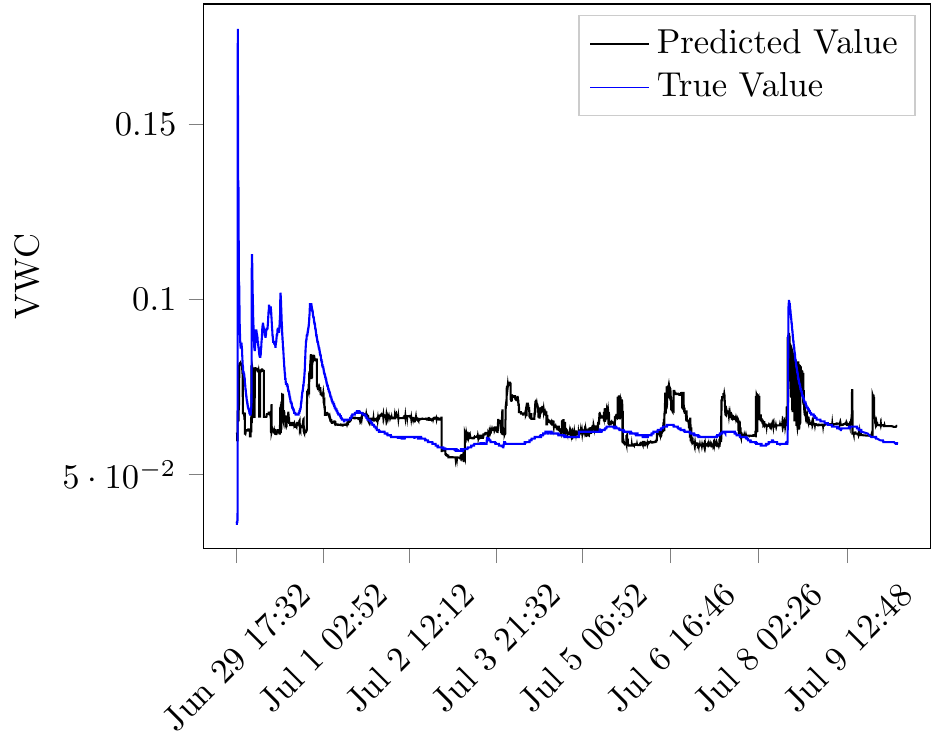}
	\end{center}
	\caption{Gradient Boosting Regression based VWC Prediction using all inputs}
\label{rfgradient_all}
\end{figure}

Figure \ref{rfgradient_all} shows the results of gradient boosting regression. The model improves the predicted values in the stable regions of the volumetric water content as compared to predictions based only on Random Forests. The pearson constant in table \ref{table:rf1gradient} also shows improvement in the Pearson's R coefficient value, hence validating the improvement in the predictor as compared to Random Forests regression modeling. However, the results in the stable region of the volumetric water content are dissimilar to the actual values of the volumetric water content making it difficult to accurately estimate the field capacity of the soil.

\begin{table}[h]
\centering
\caption{Performance indicators for Gradient Boosting Regression based VWC prediction}
\label{table:rf1gradient}
\begin{tabular}{|l|l|}
\hline
\textbf{Sensor Cost}    & 75.7€                                                                                                                         \\ \hline
\textbf{Algorithm}      & Gradient Boosting Regression                                                                                                                 \\ \hline
\textbf{Training Input} & \begin{tabular}[c]{@{}l@{}}DS18S20 and SHT10 (Temperature), \\ YL-69 and SEN13322 (Moisture),\\ SHT10 (Humidity)\end{tabular} \\
\hline
\textbf{RMSE}           & 0.00860634                                                                                                                    \\ \hline
\textbf{MAE}            & 0.00559653
 \\ \hline
\textbf{Pearson's R}    & 0.53127975
 \\ \hline
\end{tabular}
\end{table}

\subsection{Artificial Neural Networks (ANNs)}

Artificial Neural Networks (ANNs) are a class of machine learning algorithms that can also be used for classification as well as regression modeling.

 In the current work, the back-propagation algorithm is used for training the ANN. A neural network is initialized with random weights and then iteratively, based on the prediction error, gradient flows backward and corrects the randomly initialized weights till the model gives predictions with acceptable accuracy. 

In current work, Keras \cite{b31} with Tensorflow \cite{b32} backend is used to train the ANN. An ANN with 9 hidden layers (higher number of layers led to insignificant increase in performance) is used for regression based model fitting. All the inputs are normalized in the same way as it is done for the Random Forests. Exponential Linear Unit (ELU) \cite{b33} is used as the activation function.

 The weights of an ANN are initialized to non zero values to break the symmetry. He normal \cite{b34} initializer is used to initialize the kernel. According to He normal initialization, the weights should be drawn from a truncated normal distribution centered on 0 with $\sigma = sqrt(2 / n)$ where $n$ is the number of input units. The advantage of using this initialization is that it ensures that all neurons in the network initially have approximately the same output distribution and this improves the rate of convergence. In the current configuration, biases are set to zero. L2 regularization is used to avoid the overfitting problem. It penalizes the squared magnitude of all parameters directly in the loss function.

   Batch Normalization \cite{b35} is applied immediately after all fully connected layers at the beginning of the training, forcing the activations throughout the network to take on a unit Gaussian distribution. Batch Normalization makes Neural Networks significantly more robust against bad initializations. The mean squared error loss function is used to quantify the error for training the model. The Adam \cite{b36} learning rate rule, which is an adaptive learning rate method, is used to further increase the convergence rate.

In the adaptive learning rate method, weights that receive higher gradients have their effective learning rate reduced, while weights that receive small updates have their effective learning rate increased. Each model is trained for 150 epochs and in batch sizes of 128 samples with overlapping samples in subsequent batches.. 
\subsubsection*{\textbf{Results using all sensors}}
\begin{figure}[h]
	\begin{center}
		\includegraphics[scale=1]{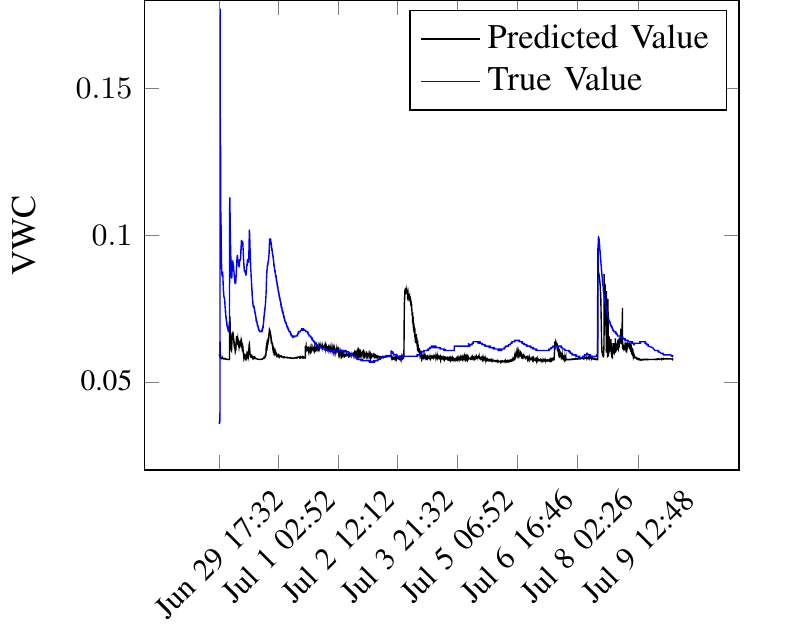}
	\end{center}
		\caption{ANN based VWC Prediction using all inputs}
\label{nn_all_111}
\end{figure}

Initially, the data collected from all the low cost sensors i.e.\ DS18S20 temperature, SHT10 humidity, SHT10 temperature, YL-69 moisture and SEN13322 moisture is used to train the ANN. Figure \ref{nn_all_111} shows  the results for the predicted value against the actual VWC value and Table \ref{table:ann1} shows the performance indicators for the same classifier. The Pearson's correlation coefficient R is 0.224, that is significantly small even compared to the regression based model generated by Random Forests. One reason can be the presence of DS18S20 and SHT10 temperature sensors data in the input because they have the least correlation with the VWC data. But the predicted value shows better correlation than Random Forests during the stable phase of the actual VWC data. Therefore field capacity can still be estimated from the predicted VWC data. The predicted VWC yields a field capacity of approximately 5.8\%.

\begin{table}[h]
\centering
\caption{Performance indicators for Neural Networks based VWC prediction}
\label{table:ann1}
\begin{tabular}{|l|l|}
\hline
\textbf{Sensor Cost}    & 75.7€                                                                                                                         \\ \hline
\textbf{Algorithm}      & Neural Networks                                                                                                               \\ \hline
\textbf{Training Input} & \begin{tabular}[c]{@{}l@{}}DS18S20 and SHT10 (Temperature), \\ YL-69 and SEN13322 (Moisture)\\ SHT10 (Humidity),\end{tabular} \\ \hline
\textbf{RMSE}           & 0.01128973                                                                                                                    \\ \hline
\textbf{MAE}            & 0.00691554                                                                                                                    \\ \hline
\textbf{Pearson's R}    & 0.22410135                                                                                                                    \\ \hline
\end{tabular}
\end{table}

\subsubsection*{\textbf{Results without temperature sensor}}

\begin{figure}[h]
	\begin{center}
		\includegraphics[scale=1]{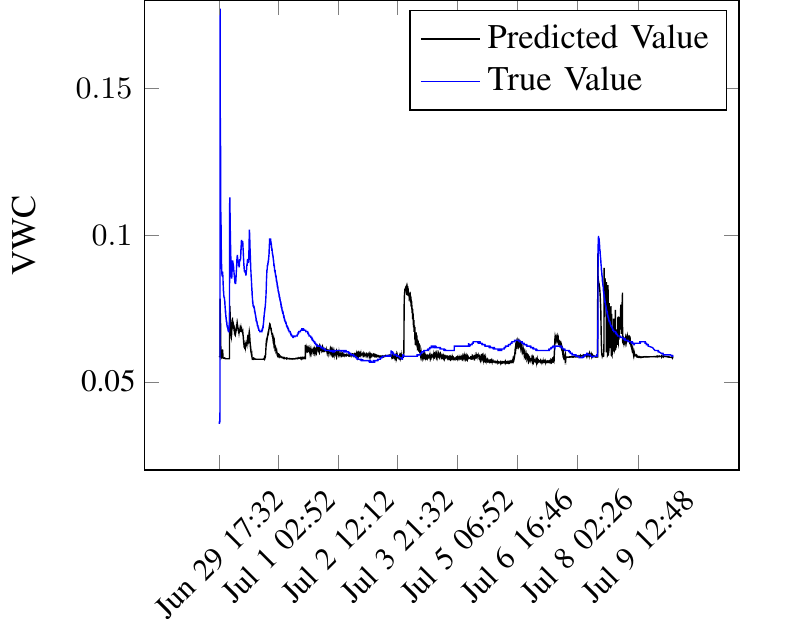}
	\end{center}
	\caption{ANN based VWC Pediction using all sensors except DS18S20}
\label{nn_nodsx}
\end{figure}

 To cross validate this, temperature sensors data is discarded while training the ANN. The data from all the other low cost sensors is used for training the ANN. Figure \ref{nn_nodsx} shows the predicted value against the actual VWC value and Table \ref{table:ann2} shows the value for different performance indicators. The Pearson correlation coefficient R increases from 0.224 to 0.4857 and the approximated field capacity is 5.7\%.

\begin{table}[h]
\centering
\caption{Performance indicators for Neural Networks based VWC prediction}
\label{table:ann2}
\begin{tabular}{|l|l|}
\hline
\textbf{Sensor Cost}    & 60.2€                                                                                                            \\ \hline
\textbf{Algorithm}      & Neural Networks                                                                                                  \\ \hline
\textbf{Training Input} & \begin{tabular}[c]{@{}l@{}}SHT10 (Temperature), \\ YL-69 and SEN13322 (Moisture)\\ SHT10 (Humidity)\end{tabular} \\ \hline
\textbf{RMSE}           & 0.01049934                                                                                                       \\ \hline
\textbf{MAE}            & 0.00650796                                                                                                       \\ \hline
\textbf{Pearson's R}    & 0.48575855                                                                                                       \\ \hline
\end{tabular}
\end{table}

\subsubsection*{\textbf{Results using only moisture sensors}}
\begin{figure}[h]
	\begin{center}
	\includegraphics{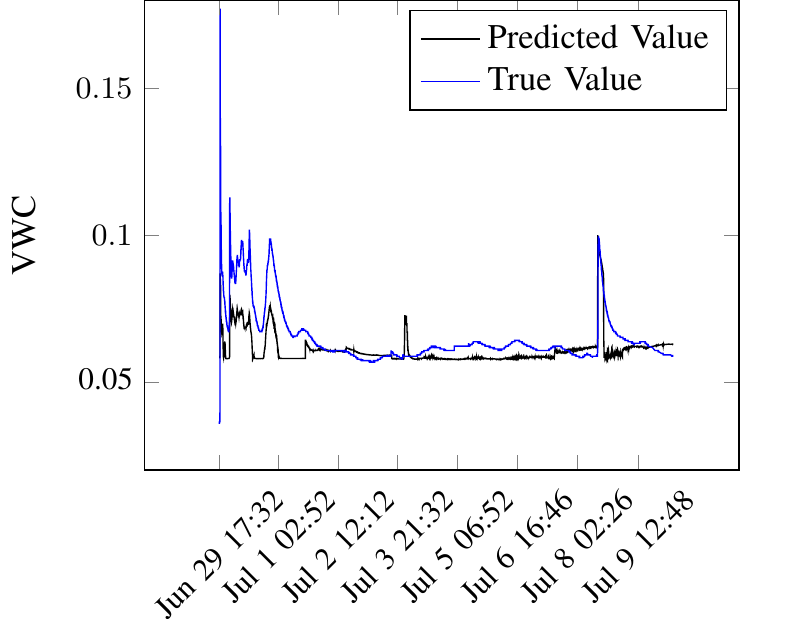}
	\end{center}
	\caption{ANN based VWC Prediction using moisture sensor YL-69 and SEN13322}
\label{nn_moisture}
\end{figure}

Finally, only the data from the moisture sensor YL-69 and SEN13322 is used as an input for training the ANN. Figure \ref{nn_moisture} shows the predicted value against the actual VWC value and the performance indicators in Table \ref{table:ann3} show significant improvement for the aforementioned configuration of inputs. The Pearson's correlation coefficient R for this configuration is equal to 0.7305 and shows maximum correlation among all the earlier configurations between the predicted and actual value for the VWC. The predicted value not only shows better correlation during the transient phase of the actual VWC data but also shows significantly less variation in the field capacity measuring region. The estimated field capacity from the predicted VWC data is approximately 5.6\%. \\
Table \ref{comparison_all} shows the summarized comparison of all the algorithms with different configuration of inputs. Regression based on ANNs with only moisture sensors not only reduces the overall cost of the sensors, but also improves the estimation accuracy of the field capacity as compared to all other configurations.

\begin{table}[h]
\centering
\caption{Performance indicators for Neural Networks based VWC prediction}
\label{table:ann3}
\begin{tabular}{|l|l|}
\hline
\textbf{Sensor Cost}    & 6.2€                          \\ \hline
\textbf{Algorithm}      & Neural Networks               \\ \hline
\textbf{Training Input} & YL-69 and SEN13322 (Moisture) \\ \hline
\textbf{RMSE}           & 0.00853104                    \\ \hline
\textbf{MAE}            & 0.00566044                    \\ \hline
\textbf{Pearson's R}    & 0.73053613                    \\ \hline
\end{tabular}
\end{table}

\section{Conclusion}
The proposed method allows much more dense deployment of the sensor nodes and hence better  field capacity profile estimation of the agricultural fields. Using low cost sensor i.e.\ YL-69 and SEN13322 moisture sensors, the VWC and field capacity of the soil can be estimated. Hence, the cost of the sensors used for soil monitoring can be significantly reduced, that can help the common farmers to use soil monitoring approaches in their agricultural fields for better crop production.  Also the current approach with extended observation datasets will help the farmers and researchers to estimate, predict and better understand the behaviour of different soils under observation.\\

\begingroup
\setlength{\tabcolsep}{6.5pt} % Default value: 6pt
\renewcommand{\arraystretch}{2} % Default value: 1

\begin{landscape}
 \topskip0pt
    \vspace*{\fill}
\begin{table}[htb!]
\centering
\begin{tabular}{|c|c|c|c|c|c|c|c|c|c|c|}
\hline
 & \textbf{\begin{tabular}[c]{@{}c@{}}DS18S20\\ (T)\end{tabular}} & \textbf{\begin{tabular}[c]{@{}c@{}}SHT10\\ (T)\end{tabular}} & \textbf{\begin{tabular}[c]{@{}c@{}}YL-69 \\ (M)\end{tabular}} & \textbf{\begin{tabular}[c]{@{}c@{}}SEN13322 \\(M)\end{tabular}} & \textbf{\begin{tabular}[c]{@{}c@{}}SHT10 \\(H)\end{tabular}} & \textbf{\begin{tabular}[c]{@{}c@{}}Sensors \\ cost (€)\end{tabular}} & \textbf{RMSE} & \textbf{MAE} & \textbf{Pearson's R} & \textbf{\begin{tabular}[c]{@{}c@{}}Field Capacity \\  Actual/Estimated\end{tabular}} \\ \hline
\textbf{\begin{tabular}[c]{@{}c@{}}Support Vector\\ Regression (SVR)\end{tabular}} & \cmark & \cmark & \cmark & \cmark & \cmark & 75.7 & 0.00927354 & 0.00564785 & 0.40901482 & Not Possible \\ \hline
\textbf{\begin{tabular}[c]{@{}c@{}}Random Forests \\Algorithm\end{tabular}} & \cmark & \cmark & \cmark & \cmark & \cmark & 75.7 & 0.00907407 & 0.00621031 & 0.47870054 & Not Possible \\ \hline
\textbf{\begin{tabular}[c]{@{}c@{}}Gradient Boosting \\Regression\end{tabular}} & \cmark & \cmark & \cmark & \cmark & \cmark & 75.7 & 0.00860634 & 0.00559653 & 0.53127975 & Not Possible \\ \hline
\textbf{\begin{tabular}[c]{@{}c@{}}Artificial Neural \\Network (ANN)\end{tabular}} & \cmark & \cmark & \cmark & \cmark & \cmark & 75.7 & 0.01128973 & 0.00691554 & 0.22410135 & 5.5\% / 5.8\% \\ \hline
\textbf{\begin{tabular}[c]{@{}c@{}}Artificial Neural \\Network (ANN)\end{tabular}} & \xmark & \cmark & \cmark & \cmark & \cmark & 60.2 & 0.01049934 & 0.00650796 & 0.48575855 & 5.5\% / 5.7\% \\ \hline
\textbf{\begin{tabular}[c]{@{}c@{}}Artificial Neural\\ Network (ANN)\end{tabular}} & \xmark & \xmark & \cmark & \cmark & \xmark & 6.2 & 0.00853104 & 0.00566044 & 0.73053613 & 5.5\% / 5.6\% \\ \hline
\end{tabular}
\caption{Performance comparison of all the algorithms (T = Temperature, M = Moisture, H = Humidity)}
\label{comparison_all}
\end{table}
    \vspace*{\fill}
\end{landscape}



\begin{thebibliography}{00}

\bibitem{b1} Challenges and Opportunities of Wireless Underground Sensor Networks, I.Zaman and A. F\"orster, GI/ITG KuVS Fachgespräch Sensornetze – FGSN 2018 

\bibitem{b2}  Ramesh, Maneesha V. "Real-Time Wireless Sensor Network for Landslide Detection", Proceedings of the 2009 Third International Conference on Sensor Technologies and Applications, SENSORCOMM '09, 2009

\bibitem{b3} J. Koskinen, P. Kilpeläinen, J. Rehu, P. Tukeva, M. Sallinen,
"Wireless Sensor Networks for infrastructure and industrial monitoring applications". ICTC 2010

\bibitem{b4} Finding Trapped Miners with Wireless Sensor Networks, I. Zaman, A. F\"orster, A. Mahmood, F. Cawood,International Conference on Information and Communication Technologies for Disaster Management 2018   
        
\bibitem{b5} Ojha, Tamoghna and Misra, Sudip and Raghuwanshi, Narendra Singh, 
"Wireless Sensor Networks for Agriculture", Comput, Electron, Agric, October 2015

\bibitem{b6} ReviTal10 Experimental field Bremen: http://www.revitec.uni-bremen.de/fotoalbum\_revi.html\#new  

\bibitem{b7} Revitec Project, An approach to combat desertification http://www.revitec.uni-bremen.de/profil.html 

\bibitem{b8} D. Entekhabi, E. G. Njoku, P. E. O’Neill et al., “The soil moisture active passive (SMAP) mission,” Proceedings of the IEEE, vol. 98, 2010.

\bibitem{b9} Leila Hassan-Esfahani, Alfonso Torres-Rua, Austin Jensen and Mac Mckee, ''Spatial root zone soil water content estimation in agricultural lands using bayesial-based artificial neural networks and high resolution visual, NIR, and thermal imagery'', Irrigation and Drainage, Volume 66, Issue 2, April 2017, Pages 273--288

 \bibitem{b10} Ceccato, P., Flasse, S., and Gregoire, J. (2002b), ''Designing a spectral index to estimate vegetation water content from remote sensing data: Part 2. Validation and applications'', Remote Sensing of Environment, 82, 198--207.
 
 \bibitem{b11} G. Shan, Y. Sun, Q. Cheng, Z. Wang, H. Zhou, L. Wang, X. Xue, B. Chen, S.B. Jones, .. ''Monitoring tomato root zone water content variation and partitioning evapotranspiration with a novel horizontally-oriented mobile dielectric sensor'', Agricultural and Forest Meteorology, November 2016
 
 \bibitem{b12} Mittelbach, H., F. Casini, I. Lehner, A. J. Teuling, and S. I. Seneviratne (2011), Soil moisture monitoring for climate research: Evaluation of a low cost sensor in the framework of the Swiss Soil Moisture Experiment (SwissSMEX) campaign, J. Geophys. Res., 116

\bibitem{b13} 5TM decagon sensor datasheet https://www.ai-nex.co.jp/5TM-Integrators-Guide.pdf   

\bibitem{b14} I. Olorunfemi, J. Fasinmirin, A. Ojo, Modeling cation exchange capacity and soil water holding capacity from basic soil properties, Eurasian J. Soil Sci. 2016

 \bibitem{b15} DS18S20 Sensor Datasheet. [Online]. Available: https://datasheets.maximintegrated.com/en/ds/DS18S20.pdf
 
 \bibitem{b16} DS18S20 Waterproof Sensor. [Online]. Available: https://www.conrad.de/de/c-control-temperatursensor-ds18s20-passend-fuer-serie-c-control-198284.html
 
 \bibitem{b17} SHT10 Sensor Datasheet. [Online]. Available: \url{https://cdn-shop.adafruit.com/datasheets/Sensirion_Humidity_SHT1x_Datasheet_V5.pdf}
 
 \bibitem{b18} SHT10 Waterproof Sensor. [Online]. Available: \url{http://www.exp-tech.de/temperatur-feuchte-sensor-sht10}

 \bibitem{b19} YL-69 Soil Moisture Sensor. [Online]. Available:https://www.fasttech.com/product/4738201-yl-69-soil-humidity-moisture-detection-sensor
 
 \bibitem{b20} SEN13322 Sensor Datasheet. [Online]. Available: https://github.com/sparkfun/Soil\_Moisture\_Sensor
 
  \bibitem{b21} SEN13322 Moisture Sensor. [Online]. Available: https://eckstein-shop.de/SparkFun-Soil-Moisture-Hygrometer-Detection-Feuchtigkeits
   
\bibitem{b23} 5TM decagon Sensor https://www.aliexpress.com/item/5TM-Soil-Moisture-and-Temperature/1866562774.html [accessed on: 27:11:2018]        
           
  \bibitem{b24} Chen R, Chen Y, Chen W, Chen Y. Time Domain Reflectometry for Water Content Measurement of Municipal Solid Waste. Environmental Engineering Science. 2012
  
  \bibitem{b25} Topp, G.C., J.L. David, and A.P. Annan 1980.  Electromagnetic, Determination of Soil Water Content:  Measurement in Coaxial Transmission Lines. Water Resources Research 

\bibitem{b27} Molenet: A new sensor node for underground monitoring, I. Zaman, J. Dede, M. Gellhaar, H. Koehler, A. F\"orster, Senseapp 2016, IEEE International Workshop on Practical Issues in Building Sensor Network Applications
  
\bibitem{b28} Support Vector Method for Function Approximation, V. Vapnik, S. Golowich and A. Smola,  Regression Estimation, and Signal Processing  Neural
Information Processing Systems, 1997
  
\bibitem{b29} Radial Basis Function Networks,Encyclopedia of Machine Learning, Buhmann, M. D, 2010   
  

 \bibitem{b30} Pedregosa, F. and Varoquaux, G. and Gramfort, A.  ''Scikit-learn: Machine Learning in Python'', Journal of Machine Learning Research, volume 12, pages 2825--2830, 2011
 
 \bibitem{b31} Keras. Chollet, Francois et. al. 2015, GitHub. https://github.com/fchollet/keras
 
 \bibitem{b32} Martín Abadi, Ashish Agarwal and Paul Barham
TensorFlow: Large-scale machine learning on heterogeneous systems, 2015. Software available from tensorflow.org.

\bibitem{b33} Djork-Arné Clevert, Thomas Unterthiner, Sepp Hochreiter, ''Fast and Accurate Deep Network Learning by Exponential Linear Units (ELUs)'', International Conference on Learning Representations 2016

\bibitem{b34} Kaiming He, Xiangyu Zhang, Shaoqing Ren, Jian Sun, ''Delving Deep into Rectifiers: Surpassing Human-Level Performance on ImageNet Classification''. Proceedings of the 2015 IEEE International Conference on Computer Vision (ICCV)

\bibitem{b35} Ioffe, Sergey and Szegedy, Christian, ''Batch Normalization: Accelerating Deep Network Training by Reducing Internal Covariate Shift'' International Conference on Machine Learning , 2015

\bibitem{b36} Diederik P. Kingma, Jimmy Ba, ''Adam: A Method for Stochastic Optimization''. International Conference on Learning Representations.


\end{thebibliography}
\end{document}